\documentclass[%
 reprint,
onecolumn,
superscriptaddress,
 amsmath,amssymb,
 aps,
 pra,
floatfix,
]{revtex4-1}

\providecommand{\linenomathWithnumbers}{}
\usepackage{newtxtext, newtxmath}
\usepackage[usenames, dvipsnames]{xcolor}
\usepackage{graphicx}
\usepackage{dcolumn} 
\usepackage{xpatch}
    \makeatletter \xpretocmd \start@align{\linenomathWithnumbers}{}{\fail}
\usepackage{mathtools, url}
\usepackage{multirow}
\usepackage{CJKutf8}
\usepackage[toc]{appendix}
\usepackage[ruled, linesnumbered]{algorithm2e}
\usepackage[unicode=true, bookmarks=true, bookmarksnumbered=false, bookmarksopen=false, breaklinks=false, pdfborder={0 0 1}, backref=false, colorlinks=false, hidelinks]{hyperref}
    \hypersetup{linkcolor=black, urlcolor=black, citecolor=black, pdfstartview={FitH}}

\usepackage{xr}
\usepackage{calc}
\usepackage{hyperref}
\usepackage{verbatim}
\usepackage{makecell}
\usepackage[ruled]{algorithm2e}
\usepackage{multirow}
\definecolor{darkgreen}{rgb}{0.0, 0.5, 0.0}
\usepackage{tikz}
\usetikzlibrary{quantikz2}
\usepackage{amssymb}
\usepackage{empheq}

\usepackage[math]{cellspace}
\DeclareSymbolFont{CMlargesymbols}{OMX}{cmex}{m}{n}
\let\sumop\relax\let\prodop\relax
\DeclareMathSymbol{\sumop}{\mathop}{CMlargesymbols}{"50}
\DeclareMathSymbol{\prodop}{\mathop}{CMlargesymbols}{"51}

\SetSymbolFont{operators}{normal}{OT1}{ntxtlf}{m}{n}
\SetSymbolFont{operators}{bold}{OT1}{ntxtlf}{b}{n}

\newcommand{\p}{\partial}
\newcommand{\dif}{\mathop{}\!\mathrm{d}}

\newcommand{\ee}{\mathrm{e}}
\newcommand{\ii}{\mathrm{i}}

\renewcommand{\le}{\leqslant}

\renewcommand{\ket}[1]{| #1 \rangle}

\newcommand{\T}{^{\mathrm{T}}}

\allowdisplaybreaks

\begin{document}
\title{Quantum algorithm for the nonlinear Schr\"odinger equation via the Lax-pair scattering} 
\author{Chenjia Zhu} 
 \affiliation{State Key Laboratory for Turbulence and Complex Systems, School of Mechanics and Engineering Science, Peking University, Beijing 100871, China}%
 \affiliation{HEDPS-CAPT, Peking University, Beijing 100871, China}
\author{Zhaoyuan Meng} 
 \affiliation{State Key Laboratory of Nonlinear Mechanics, Institute of Mechanics, Chinese Academy of Sciences, Beijing 100190, PR China}
\author{Yousheng Zhang} 
 \affiliation{Institute of Applied Physics and Computational Mathematics, Beijing, 100094, China}
 \affiliation{HEDPS-CAPT, Peking University, Beijing 100871, China}
 \affiliation{National Key Laboratory of Computational Physics, Beijing, 100088, China}
\author{Yue Yang} 
 \email{yyg@pku.edu.cn}
 \affiliation{State Key Laboratory for Turbulence and Complex Systems, School of Mechanics and Engineering Science, Peking University, Beijing 100871, China}%
 \affiliation{HEDPS-CAPT, Peking University, Beijing 100871, China}

\date{\today}

\begin{abstract}
The nonlinear Schr\"odinger equation (NLSE) governs a broad class of wave phenomena, including deep-water waves, quantum turbulence, and solitons.
The multiscale spatiotemporal coupling inherent in these systems imposes severe computational bottlenecks on classical high-fidelity numerical simulations.
While quantum computing offers the potential for exponential speedup, its unitary dynamics pose a fundamental challenge to solve the NLSE. 
We propose a quantum framework based on the Lax-pair scattering for solving the 1D NLSE.
Specifically, the physical field is first mapped into the spectral space via a quantum direct scattering circuit.
Following a decoupled linear time evolution, the physical solution is reconstructed through an inverse scattering transform utilizing the quantum singular value transformation.
Since the temporal evolution is performed analytically in the scattering domain, the framework bypasses iterative time stepping, rendering it highly advantageous for long-time simulations.
To demonstrate the accuracy and noise resilience of this approach, we simulate a Gaussian wave packet under quantum noise, two-soliton collisions, breather dynamics, and modulational instability on a quantum emulator. 
\end{abstract}

\maketitle

\section{Introduction}
The nonlinear Schr\"odinger equation (NLSE) is a cornerstone of nonlinear physics, serving as a universal model for wave propagation in weakly nonlinear and dispersive media~\cite{sulem2007nonlinear, liu2019schrodinger}.
It describes a wide range of physical phenomena across disparate length and energy scales, including rogue-wave formation in deep-water hydrodynamics~\cite{onorato2001freak, osborne2001random,osborne2019breather}, pulse propagation in nonlinear optical fibers~\cite{barthelemy1985propagation, solli2007optical, bonatto2011deterministic}, wave-plasma interactions in fusion physics~\cite{rudakov1978strong, chen2024drift}, Bose-Einstein condensates~\cite{gross1961structure, pitaevskii1961vortex,rogel2013gross, takeuchi2010quantum}, and quantum turbulence~\cite{nazarenko2019wave, kobayashi2021quantum, ferrini2025driven}.  

Despite its simple form, the NLSE exhibits rich nonlinear dynamics, including solitons, breathers, rogue waves, and wave turbulence~\cite{dudley2014instabilities, osborne2019breather}. 
Interactions among these coherent structures drive broadband spectral transfer, rapid phase variation, and the localization of wave packets across disparate spatial and temporal scales. 
Such dynamics are particularly prominent in quantum turbulence~\cite{nazarenko2019wave, ferrini2025driven} and deep-water waves~\cite{osborne2019breather}. 
Consequently, accurate long-time simulations demand numerical schemes with high spatial resolution, long-time stability, and robust conservation properties~\cite{antoine2013computational}. 
The computational cost of classical simulations scales rapidly with the required resolution and evolution time, particularly in turbulent or higher-dimensional regimes.

Quantum computing offers a promising paradigm to accelerate large-scale scientific computation by exploiting quantum superposition and entanglement~\cite{Nielsen2010Quantum}.
Prominent examples include Shor's factoring algorithm~\cite{Shor1999polynomial}, Grover's search algorithm~\cite{grover1996fast}, quantum linear system solvers~\cite{harrow2009quantum, childs2017quantum}, and quantum-inspired methods~\cite{gourianov2022quantum}.
For linear equations, systematic frameworks such as Schr\"odingerization~\cite{jin2023quantum, jin2024quantum, jin2025schrodingerization} and the linear combination of Hamiltonian simulation~\cite{an2023linear, an2026quantum, yang2025circuit, Meng2026_Toward} have been developed.
However, nonlinear partial differential equations pose a formidable challenge.
While standard quantum circuits implement linear unitary operations, nonlinear dynamics do not directly conform to this computational paradigm~\cite{tennie2025quantum}.
A central objective is therefore to reformulate nonlinear evolution in a quantum-compatible form without introducing prohibitive overhead.

Existing approaches to bridging this gap generally fall into two categories.
The first category reformulates the nonlinear dynamics by embedding the original system in high-dimensional linear spaces amenable to quantum evolution, such as Carleman linearization~\cite{liu2021efficient, itani2022analysis, sanavio2024three}, the Koopman operator~\cite{giannakis2022embedding, zhang2025data, gan2025provably}, the homotopy analysis method~\cite{liao2024ham, xue2025quantum}, quantum spin encoding~\cite{meng2023quantum, meng2024quantum, meng2024simulating, su2026quantum}, and lattice gas cellular automata~\cite{yepez2001quantum, Wang2025_Quantum}. 
While these approaches provide systematic pathways to quantum representations of nonlinear dynamics, they typically suffer from rapidly expanding auxiliary dimensions, significant truncation overhead, and accumulation errors as the nonlinear complexity increases.

The second category relies on hybrid quantum-classical frameworks.
Instead of employing explicit higher-dimensional embeddings, these methods approximate nonlinear dynamics through classical feedback or variational optimization.
Representative examples include quantum variational solvers~\cite{lubasch2020variational, leong2022variational, fathi2024hybrid} and hybrid algorithms in which quantum subroutines alleviate linear-algebra bottlenecks~\cite{bharadwaj2023hybrid, ye2024hybrid, gnanasekaran2024variational}.
Recent quantum algorithms for the NLSE include operator-splitting schemes in which the quantum Fourier transform implements the linear evolution, while either parameterized quantum circuits~\cite{kocher2025numerical} or measurement-assisted potential reconstruction~\cite{weng2026quantum} resolves the nonlinear term.
An alternative approach targets the NLSE with a Ginzburg-Landau potential~\cite{jin2026hybrid} by coupling classical vortex dynamics with quantum linear-system solvers for the associated elliptic equations.
Despite their flexibility, these hybrid approaches typically suffer from accumulated time stepping errors, recurrent measurement overhead, optimization instabilities, and a formulation-dependent loss of long-time accuracy.
These limitations ultimately restrict their scalability and reliability for long-time simulations.

Developing quantum algorithms for high-dimensional nonlinear systems, such as quantum turbulence, requires a tractable starting point that captures the essential features of nonlinear wave dynamics.
The 1D NLSE serves as a prototypical framework for this purpose.
As a paradigmatic integrable system, it admits a Lax pair formulation~\cite{lax1968integrals, shabat1972exact, ablowitz1974inverse}.
This formulation maps the nonlinear physical-space dynamics to a linear, decoupled evolution in an auxiliary spectral space, from which the physical solution is reconstructed via an inverse transform.
The Lax framework thereby avoids the repeated nonlinear time stepping required by most existing quantum simulation methods, reducing the evolution to a linear spectral update.
This structure makes the 1D NLSE a natural starting point for quantum algorithmic design.
Moreover, under suitable data-access assumptions, both the forward and inverse scattering transforms admit efficient quantum subroutines.
Together, these properties provide a systematic pathway to the quantum simulation of the 1D NLSE.

\begin{figure*}
    \centering
    \includegraphics[width=\linewidth]{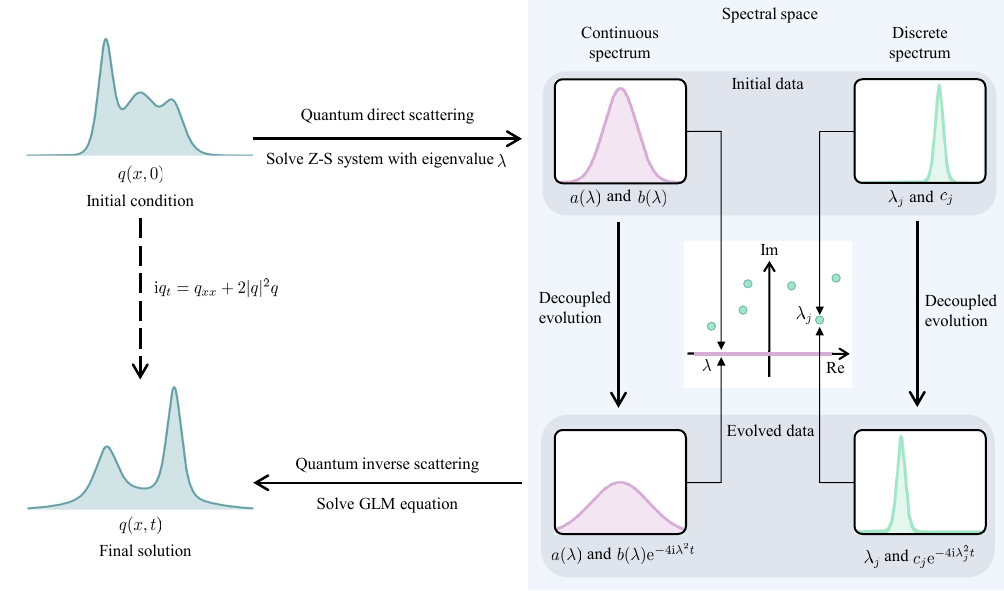}
    \caption{Overview of the quantum algorithm for the 1D NLSE.
    The algorithm starts with a quantum direct scattering transform, which maps the initial field $q(x,0)$ to the spectral domain by solving the Zakharov-Shabat (Z-S) system.
    Here, the eigenvalue $\lambda$ serves as the spectral parameter of the transform.
    The initial wave function is decomposed into its intrinsic spectral components, consisting of the continuous spectrum with $\lambda \in \mathbb{R}$ for dispersive radiation and the discrete spectrum with $\lambda_j \in \mathbb{C}$ for soliton dynamics.
    These two spectral components then evolve independently in time.
    The evolved field $q(x,t)$ is finally reconstructed through quantum inverse scattering by solving the Gelfand-Levitan-Marchenko (GLM) integral equation with quantum singular value transformation. 
    }
    \label{fig:overview}
\end{figure*}

Leveraging the Lax framework, we propose a quantum algorithm for solving the 1D NLSE.
As illustrated in Fig.~\ref{fig:overview}, the algorithm consists of three primary steps.
First, we construct a quantum circuit for the Zakharov-Shabat (Z-S) system, which maps the physical field to the spectral data via a quantum direct scattering transform.
The scattering data, comprising the continuous and discrete spectra, then undergo independent linear evolution in spectral space.
Finally, the evolved physical field is reconstructed through a quantum inverse transform based on the Gelfand-Levitan-Marchenko (GLM) equation.
We validate the accuracy of this framework by analyzing several representative physical scenarios, including Gaussian wave packets in the presence of quantum noise, soliton collisions, breather dynamics, and modulational instability.
The numerical results demonstrate well-controlled errors and strong agreement with classical solutions.

The remainder of this paper is organized as follows.
In Sec.~\ref{sec:math}, we introduce the NLSE, its Lax pair formulation, and the Lax method for solving its dynamics.
In Sec.~\ref{sec:quantum}, we detail the quantum algorithms for the three primary procedures of the proposed framework, followed by a computational complexity analysis.
We analyze the numerical simulation results in Sec.~\ref{sec:result}.
Finally, concluding remarks are presented in Sec.~\ref{sec:conclusion}. 

\section{Mathematical framework}
\label{sec:math}
\subsection{Lax pairs of 1D NLSE} \label{sec:laxpair}

We briefly introduce the Lax-pair method for the 1D NLSE~\cite{lax1968integrals}, which provides a spectral formulation of the integrable dynamics.
This formulation utilizes a time-independent spectral parameter $\lambda$ to characterize localized structures, thereby mapping coupled nonlinear dynamics into the decoupled evolution of these fundamental components.
The associated Lax method is conceptually regarded as a nonlinear analogue of the Fourier transform. 
By introducing a pair of auxiliary linear operators known as the Lax pair, this formulation maps nonlinear dynamics in the physical space into a decoupled linear evolution in the spectral space. 

Within this framework, the 1D NLSE 
\begin{equation}\label{eq:nls}
    \ii \frac{\partial q(x, t)}{\partial t} = \frac{\partial^2 q(x, t)}{\partial x^2} + 2|q(x, t)|^2 q(x, t)
\end{equation}
is reformulated as the zero-curvature condition~\cite{lax1968integrals, shabat1972exact}
\begin{equation}
    U_t - V_x + \left(UV -VU\right) = 0,
    \label{eq:zerocurvature}
\end{equation}
associated with the spatial operator 
\begin{equation}
    U =  \begin{pmatrix} -\ii \lambda & q \\ -q^* & \ii \lambda \end{pmatrix}
\end{equation}
and the temporal operator
\begin{equation}
    V = \begin{pmatrix}
        2\ii\lambda^2 - \ii |q|^2 & -2\lambda q - \ii q_x \\
        2\lambda q^* - \ii q^*_x  & -2\ii\lambda^2 + \ii |q|^2
    \end{pmatrix},
\end{equation}
where $U$ and $V$ constitute the Lax pair, and the superscript ``*'' denotes the complex conjugate.

Then, an auxiliary two-component wave function $\psi(x, t, \lambda)$ is introduced, leading to a pair of linear equations
\begin{subequations}
    \begin{empheq}[left=\empheqlbrace]{align}
        \p_x\psi &= U \psi, \label{eq:laxspace} \\
        \p_t\psi &= V \psi. \label{eq:laxtime}
    \end{empheq}
\end{subequations}
The zero-curvature condition in Eq.~\eqref{eq:zerocurvature} ensures that Eqs.~\eqref{eq:laxspace} and \eqref{eq:laxtime} are compatible when Eq.~\eqref{eq:nls} is satisfied. 
Specifically, $\psi$ satisfying Eq.~\eqref{eq:laxspace} at $t=0$ remains a solution to Eq.~\eqref{eq:laxspace} at any subsequent time $t$ under the evolution governed by Eq.~\eqref{eq:laxtime}.
Moreover, Eq.~(\ref{eq:zerocurvature}) yields $\partial \lambda/\partial t = 0$, identifying $\lambda$ as an invariant of the 1D NLSE. 

Consequently, this compatibility and the invariance of $\lambda$ are exploited to decouple the system dynamics.
Here, Eq.~\eqref{eq:laxspace} serves as the transformation mapping the physical field into the spectral space, and  
Eq.~\eqref{eq:laxtime} governs the temporal evolution exclusively within this spectral space.

Using the Lax pair formulation, we solve the initial value problem of Eq.~\eqref{eq:nls} through three steps. 
First, the direct scattering transform involves the spatial Eq.~\eqref{eq:laxspace} at $t=0$, which maps the physical field to the scattering data in the spectral space. 
Subsequently, these scattering data evolve according to the temporal relationship derived from Eq.~\eqref{eq:laxtime}. 
Finally, the inverse scattering transform maps the evolved spectral data back to the physical space to reconstruct the desired solution $q(x,t)$. 


\subsection{Asymptotic behavior of Lax pairs}
The asymptotic analysis of Eqs.~\eqref{eq:laxspace} and \eqref{eq:laxtime} facilitates their solution.
Direct integration of Eq.~\eqref{eq:laxtime} remains intractable due to its explicit dependence on the unknown evolving field $q(x,t)$.
However, for physical fields that decay rapidly as $|x| \to \infty$, the terms involving $q$ vanish, and the Lax operators in Eqs.~\eqref{eq:laxspace} and \eqref{eq:laxtime} simplify to constant-coefficient matrices.
This asymptotic simplification yields a tractable framework for solving the equations.
Consequently, instead of solving the coupled system directly, we investigate the asymptotic behavior of both operators at spatial infinity.

We begin with the spatial spectral equation Eq.~\eqref{eq:laxspace}, reformulating it into the standard Z-S system~\cite{shabat1972exact, ablowitz1974inverse} 
\begin{equation}
    - \ii \begin{pmatrix} 1 & 0 \\ 0 & -1 \end{pmatrix} \frac{\partial \psi}{\partial x} - \ii \begin{pmatrix} 0 & q(x, 0) \\ q^{*}(x, 0) & 0 \end{pmatrix} \psi = \lambda \psi.
    \label{eq:zssysstem}
\end{equation}
In the limit $q(x, 0) \to 0$ as $|x| \to \infty$, the Z-S system reduces to a constant-coefficient system.
This property uniquely defines two fundamental matrix solutions, the left and right Jost solutions $\varPsi_l$ and $\varPsi_r$~\cite{ablowitz1974inverse, arico2011numerical}, which independently match the asymptotic boundary conditions at $x \to \pm\infty$.
These $2 \times 2$ matrices are expressed as
\begin{equation}
    \varPsi_l = \begin{pmatrix}
        \psi_{l, 1}^{(1)} & \psi_{l, 2}^{(1)} \\
        \psi_{l, 1}^{(2)} & \psi_{l, 2}^{(2)} 
    \end{pmatrix}
    = (\psi_{l,1}, \psi_{l,2}) \quad \text{and} \quad
    \varPsi_r = \begin{pmatrix}
        \psi_{r, 1}^{(1)} & \psi_{r, 2}^{(1)} \\
        \psi_{r, 1}^{(2)} & \psi_{r, 2}^{(2)} 
    \end{pmatrix}
    = (\psi_{r,1}, \psi_{r,2}),
\end{equation}
where the individual columns $\psi_{\alpha,j}=(\psi_{\alpha, j}^{(1)}, \psi_{\alpha, j}^{(2)})\T$ for $\alpha \in \{l,r\}$ and $j \in \{1,2\}$ constitute a basis for the solution space $\mathbb{C}^2$.
Based on the asymptotic behavior of Eq.~\eqref{eq:zssysstem}, the far-field limits of the Jost solutions are given by
\begin{subequations}
\begin{align}
\varPsi_l(\lambda, x) &= \exp(\ii \lambda x \sigma_z), \quad x \to +\infty, \label{eq:jostleft} \\
\varPsi_r(\lambda, x) &= \exp(\ii \lambda x \sigma_z), \quad x \to -\infty, \label{eq:jostright}
\end{align}
\label{eq:jostSolution}
\end{subequations}
with the Pauli-Z matrix $\sigma_z$.

Since $\psi_l$ and $\psi_r$ are fundamental solution matrices of the same Z-S system, they are linearly related by
\begin{equation}
    \varPsi_r(\lambda, x) = \varPsi_l(\lambda, x) \begin{pmatrix} a^*(\lambda) & -b(\lambda) \\ b^*(\lambda) & a(\lambda) \end{pmatrix},
    \label{eq:scatterMatrix}
\end{equation}
where $a(\lambda)$ and $b(\lambda)$ are the scattering coefficients satisfying $|a(\lambda)|^2 + |b(\lambda)|^2=1$, and $((a^*(\lambda), b^*(\lambda))\T, (-b(\lambda), a(\lambda))\T)$ is the scattering matrix~\cite{ablowitz1974inverse, arico2011numerical}.
For each spectral parameter $\lambda$, the potential $q(x,0)$ couples the components of the Z-S system during propagation across the spatial domain.
Accordingly, $a(\lambda)$ and $b(\lambda)$ characterize the transmission- and reflection-like responses induced by the entire potential profile.
Consequently, the spatial profile of the initial field $q(x,0)$ is globally encoded within the spectral scattering coefficients $a(\lambda)$ and $b(\lambda)$.

The time-independent spectral parameter $\lambda$ characterizes independent structures in the evolution.
The eigenvalue analysis of Eq.~\eqref{eq:zssysstem} yields two distinct types of $\lambda$: continuous real $\lambda$ and discrete complex $\lambda$ with $\operatorname{Im}(\lambda) > 0$~\cite{ablowitz1974inverse}. 
The real eigenvalues $\lambda \in \mathbb{R}$, also referred to as continuous spectrum, correspond to the dispersive radiation of the system, where the scattering coefficients $a(\lambda)$ and $b(\lambda)$ are complex functions defined along the real $\lambda$-axis.
The complex eigenvalues with $\operatorname{Im}(\lambda) > 0$, known as discrete spectrum, correspond to the solitons of the system, where each discrete eigenvalue $\lambda_j$ determines the amplitude and velocity of the respective soliton.
These discrete eigenvalues are also the zeros of $a(\lambda)$, at which the Jost solutions become linearly dependent.
This linear dependence intrinsically defines a normalization constant $c_j = -\ii b(\lambda_j)/a'(\lambda_j)$, which is an essential parameter for spatial reconstruction that determines the soliton's initial position and phase.

Substituting Eqs.~\eqref{eq:jostSolution} and \eqref{eq:scatterMatrix} into Eq.~\eqref{eq:laxtime} yields
\begin{equation}
    a(\lambda) \to a(\lambda), \quad b(\lambda) \to b(\lambda)\ee^{-4\ii\lambda^2 t}, \quad c_j \to c_j \ee^{-4\ii\lambda_j^2 t}, 
    \label{eq:timeEvolution}
\end{equation}
revealing that the time evolution of the scattering data is decoupled and linear.
Consequently, the mapping from the physical field to the scattering data $\{a(\lambda), b(\lambda), c_j \}$ linearizes the nonlinear evolution and facilitates solving the NLSE. 

\subsection{Direct scattering transform}
We further detail the procedure for obtaining the scattering data from Eq.~(\ref{eq:laxspace}) via the direct scattering transform, which involves two primary parts. For the continuous spectrum, the scattering data are obtained by integrating Eq.~\eqref{eq:zssysstem} and matching the asymptotic behaviors of the Jost solutions at spatial infinity. For the discrete spectrum, the eigenvalues $\lambda_j$ are extracted from the Z-S system, and the Wronskian is then utilized to express $a(\lambda)$ for the calculation of the constants $c_j = -\ii b(\lambda_j) / a'(\lambda_j)$~\cite{ablowitz1974inverse}.

First, we consider the continuous spectrum by examining the constituent column vectors of the Jost solutions. 
Extracting the first column of the matrix scattering relation in Eq.~(\ref{eq:scatterMatrix}) yields the vector equation
\begin{equation}
    \psi_{r,1}(\lambda, x) = a^*(\lambda) \psi_{l,1}(\lambda, x) + b^*(\lambda) \psi_{l,2}(\lambda, x).
\end{equation}
According to the asymptotic condition in Eq.~\eqref{eq:jostright}, we obtain $\psi_{r,1}(\lambda, x) = (\ee^{\ii\lambda x}, 0)\T$ at $x \to - \infty$.
The asymptotic behavior of $\psi_l$ in Eq.~\eqref{eq:jostleft} implies $\psi_{l,1} = (\ee^{\ii\lambda x}, 0)\T$ and $\psi_{l,2} = (0, \ee^{-\ii\lambda x})\T$ in the limit $x \to +\infty$.
Comparing the left and right Jost solutions in this asymptotic regime yields 
\begin{equation} 
    \psi_{r,1}(\lambda, x) = \begin{pmatrix} a^*(\lambda) \ee^{\ii\lambda x} \\ b^*(\lambda) \ee^{-\ii\lambda x} \end{pmatrix}, \quad x \to +\infty.
        \label{eq:asymptoticMatch}
\end{equation}
Based on this asymptotic matching, $a(\lambda) = \lim_{x \to +\infty} \psi^{(1)*}_{r, 1}(\lambda, x) \ee^{\ii\lambda x}$ and $b(\lambda) = \lim_{x \to +\infty} \psi^{(2)*}_{r, 1}(\lambda, x) \ee^{-\ii\lambda x}$ are obtained from the components of $\psi_{r, 1}$ by taking the appropriate limits. 
Here, 
\begin{equation}
    \psi_{r, 1}(\lambda, x) = \mathcal{P} \exp \left( \int_{x_0}^{x} \begin{pmatrix} \ii \lambda & -q(s,0) \\ q^*(s,0) & -\ii \lambda \end{pmatrix} \mathrm{d}s \right) \begin{pmatrix} \ee^{\ii \lambda x_0} \\ 0 \end{pmatrix},~x_0 \to -\infty
    \label{eq:zsIntegral}
\end{equation}
is obtained by integrating Eq.~\eqref{eq:zssysstem}, where $\mathcal{P}$ denotes the path-ordering operator, which ensures that matrices evaluated at larger values of $s$ appear to the left in the expansion.

Consequently, the extraction of continuous scattering data $\{a(\lambda), b(\lambda)\}$ requires numerical spatial integration in Eq.~\eqref{eq:zsIntegral}, typically performed via the transfer matrix approach~\cite{boffetta1992computation} or solving integral equations.
The determination of $a(\lambda)$ and $b(\lambda)$ for all $\lambda \in \mathbb{R}$ necessitates the repeated numerical evaluation of this integration over a finely discretized spatial grid for all $\lambda_i \in \mathbb{R}$, which constitutes a significant computational bottleneck.

We next consider the discrete spectrum.
The discrete eigenvalues $\lambda_j$ are obtained by solving the eigenvalue problem associated with the Z-S system in Eq.~\eqref{eq:zssysstem}.
To evaluate the corresponding normalization coefficient $c_j = -\ii b(\lambda_j) / a'(\lambda_j)$, we first express the scattering coefficient $a(\lambda)$ in terms of the Wronskian
\begin{equation}
    a(\lambda) =  W(\psi_{l, 1}(\lambda, x), \psi_{r, 2}(\lambda, x)).
    \label{eq:Wronskian}
\end{equation}
Following the derivations detailed in Appendix~\ref{sec:coeffDerivation}, we obtain
\begin{equation}
    a'(\lambda_j) = -2\ii b(\lambda_j) \int_{-\infty}^{\infty} \psi_{l,1}^{(1)}(\lambda_j, s) \psi_{l,1}^{(2)}(\lambda_j, s) \mathrm{d}s,
    \label{eq:aDiff}
\end{equation}
which subsequently yields the normalization coefficient
\begin{equation}
    c_j  = \frac{1}{2  \int_{-\infty}^{\infty} \psi_{l,1}^{(1)}(\lambda_j, s) \psi_{l,1}^{(2)}(\lambda_j, s) \mathrm{d}s }.
    \label{eq:coefficient}
\end{equation}

\subsection{Inverse scattering transform}
The procedure concludes by reconstructing the physical field $q(x,t)$ from the evolved scattering data.
The inverse scattering transform requires solving the GLM integral equation~\cite{gasymov1966inverse, gasymov1966determination}
\begin{equation}
    B(x, y) + \int_0^\infty \Omega(y+z+2x) B(x, z) \mathrm{d} z = H(x, y), \quad y \geq 0,
    \label{eq:glm}
\end{equation}
with 
\begin{equation}
    B(x, y) = \begin{pmatrix} B_1(x, y) \\ B_2(x, y) \end{pmatrix}, \quad 
    \Omega(s) = \begin{pmatrix} 0 & -\omega(s) \\ \omega^*(s) & 0 \end{pmatrix}, \quad 
    H(x, y) = \begin{pmatrix} 0 \\ -\omega^*(y+2x) \end{pmatrix}.
\end{equation}
Here, $B_1$ and $B_2$ are unknowns, $\Omega$ is the integral kernel determined by the time-evolved scattering data, and the scalar integral kernel
\begin{equation}
    \omega(y) = -\frac{1}{2\pi} \int_{-\infty}^{+\infty} \frac{b^*(\lambda)}{a(\lambda)} \ee^{\ii\lambda y} \mathrm{d}\lambda + \sum_{j=1}^{N_b} c_j \ee^{\ii\lambda_j y}
\end{equation}
encompasses both the continuous spectrum and the discrete spectrum, where $N_b$ denotes the number of discrete eigenvalues.
Following the solution of Eq.~\eqref{eq:glm}, $q(x, t) = 2 B_2(x, 0)$ is recovered. 

Despite its mathematical elegance, the implementation of this procedure on a classical computer encounters severe computational bottlenecks, particularly in the high-precision extraction of scattering data and the inversion of large dense matrices arising from the discretized GLM equation.
By assigning the computationally intensive direct and inverse scattering transforms to specialized quantum algorithms, the proposed method mitigates these bottlenecks through quantum parallelism.

\section{Quantum algorithms}
\label{sec:quantum}
\subsection{Direct scattering transformation} \label{sec:quantumDirect} 
We design a quantum direct scattering circuit to map $q(x,0)$ into the spectral space.
The spatial differential operator of the Z-S system in Eq.~\eqref{eq:zssysstem} naturally serves as a generator of unitary dynamics.
By treating the spatial coordinate $x$ as a pseudo-time variable and encoding the spectral parameter $\lambda$ into an auxiliary quantum register, we leverage quantum parallelism to simultaneously evolve the system across all discretized real values of $\lambda$.

We first consider the evolution for a fixed $\lambda$ in a quantum circuit.
The spatial evolution in Eq.~\eqref{eq:zssysstem} is mapped to a Schr\"odinger equation $\ii\p_x\psi=H\psi$ governed by the Hamiltonian
\begin{equation}
    H(x)= -\lambda \sigma_z + \mathrm{Im}[q(x,0)]\sigma_x + \mathrm{Re}[q(x,0)]\sigma_y,
\end{equation}
where $\sigma_\alpha$, $\alpha\in\{x,y,z\}$ denotes the Pauli matrices.
Within a quantum circuit, this evolution is conducted on a single-qubit register, hereafter denoted as a target register $T$.
The diagonal component $\lambda \sigma_z$ is entirely determined by the spectral parameter, whereas the off-diagonal component $H_{\mathrm{od}}(x) = \mathrm{Im}[q(x,0)]\sigma_x + \mathrm{Re}[q(x,0)]\sigma_y$ depends exclusively on $q(x)$.

For a small spatial step $\delta_x$, we apply the first-order Lie-Trotter decomposition~\cite{trotter1959product, lloyd1996universal} to approximate the evolution operator
\begin{equation}
    \ee^{-\ii H(x) \delta_x} = \ee^{\ii \lambda  \delta_x \sigma_z} \ee^{-\ii \mathrm{Im}[q(x,0)]\delta_x \sigma_x } \ee^{-\ii \mathrm{Re}[q(x,0)]\delta_x \sigma_y } + \mathcal{O}(\delta_x^2)
\end{equation}
as a product of independent terms. 
The off-diagonal components are implemented via the rotation gates $R_x(2\text{Im}[q(x,0)] \delta_x)$ and $R_y(2\text{Re}[q(x,0)]\delta_x)$, respectively, while the diagonal term is realized by a standard $R_z(-2\lambda\delta_x)$ rotation.

\begin{figure*}
\centering
\begin{quantikz}[row sep=0.4cm, column sep=0.25cm]
    \lstick{$|T\rangle$} & \gate{e^{-\ii H_{\mathrm{od}}(x_j) \delta_x}} & \gate{R_z(-2\delta_\lambda \delta_x)} & \gate{R_z(-4\delta_\lambda \delta_x)} & \qw ~~\dots~~ & \gate{R_z(-2^{m} \delta_\lambda \delta_x)} & \gate{R_z(-2\lambda_{\min} \delta_x)} & \qw \\
    \lstick[4]{$|A\rangle$} 
    & \qw & \ctrl{-1} & \qw & \qw ~~\dots~~ & \qw & \qw & \qw \\
    & \qw & \qw & \ctrl{-2} & \qw ~~\dots~~ & \qw & \qw & \qw \\[-0.2cm]
    \setwiretype{n} & & & & \vdots & & & \\[-0.2cm]
    & \qw & \qw & \qw & \qw ~~\dots~~ & \ctrl{-4} & \qw & \qw
    \end{quantikz}
    \caption{Quantum circuit for the single-step spatial evolution operator $U(x_i)$ in Eq.~\eqref{eq:U_xi}.
    The single-qubit target register $|T\rangle$ first undergoes an unconditional rotation $\exp(-\ii H_{\mathrm{od}}(x_i) \delta_x)$ determined by the scattering potential.
    A sequence of controlled-$R_z$ gates then acts on $|T\rangle$, with each gate selected by a computational basis state of the $m$-qubit ancillary register $|A\rangle$.
    These controlled rotations accumulate the fractional increments of the spectral phase.
    The final unconditional $R_z$ gate incorporates the negative spectral offset set by $\lambda_{\min}$.}
    \label{fig:circuitUs}
\end{figure*}
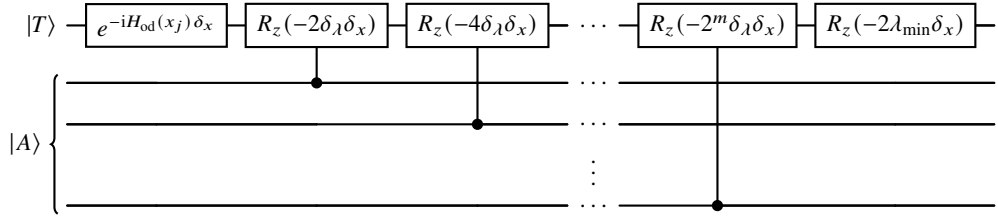

Next, we introduce a $m$-qubit ancillary register $A$ to compute the evolution across all values of $\lambda$ simultaneously.
To represent a discretized grid of $2^m$ spectral points, we define an affine mapping 
\begin{equation}\label{eq:affine_map}
    \lambda_k = \lambda_{\min} + k \delta_\lambda, k \in \{0,1,2,\cdots, 2^m-1\}
\end{equation}
that accommodates both negative offsets and fractional resolutions. 
By applying this mapping, the $Z$-rotation of the target qubit is transformed into a sequence of controlled operations governed by the computational basis $|n\rangle_A$ of the ancillary register.
Consequently, the unified single-step evolution operator acting on the composite space $A \otimes T$ becomes
\begin{equation}\label{eq:U_xi}
    U(x_i) \approx \left[ \left( I_A \otimes \ee^{\ii \lambda_{\min} \delta_x \hat{\sigma}_z} \right) \prod_{k=0}^{m-1} \Lambda_k \big( R_z(-2^{k+1} \delta_\lambda \delta_x) \big) \right] \left( I_A \otimes \ee^{-\ii H_{\mathrm{od}}(x_i) \delta_x} \right),
\end{equation}
where $\Lambda_k(V) \equiv |0\rangle\langle 0|_k \otimes I_T + |1\rangle\langle 1|_k \otimes V$ denotes a unitary operation $V$ controlled by the $k$-th ancillary qubit, and $\ee^{-\ii H_{\mathrm{od}}(x_j) \delta_x}$ represents the off-diagonal components for simplicity.
The quantum circuit that implements this operator is shown in Fig.~\ref{fig:circuitUs}.

\begin{figure*}
    \centering
    \begin{quantikz}[row sep=0.4cm, column sep=0.5cm]
    \lstick{$|0\rangle_T$} & \qw &\qw & \gate[5]{U(x_0)}  & \qw ~~\dots~~ & \gate[5]{U(x_{N - 1})} & \meter{} \\
    \lstick[4]{$|0\rangle_A^{\otimes m}$} 
    & \gate{H} & \gate{P(\delta_\lambda x_0)} & \qw & \qw ~~\dots~~ & \qw & \meter{} \\
    & \gate{H} & \gate{P(2\delta_\lambda x_0)} & & \qw ~~\dots~~ & & \meter{} \\[-0.2cm]
    \setwiretype{n} & \vdots & \vdots & & \vdots & & \vdots \\[-0.1cm]
    & \gate{H} & \gate{P(2^{m-1}\delta_\lambda x_0)} & & \qw ~~\dots~~ & & \meter{}
    \end{quantikz}
    \caption{Quantum circuit for extracting the scattering data over the discretized spectrum of $\lambda$.
    The initial boundary conditions are encoded by local single-qubit phase gates denoted by $P$.
    The spatial domain is discretized into $N$ grid points.
    At each grid point $x_i$, the unitary operator $U(x_i)$ implements the corresponding single-step evolution, as detailed in Fig.~\ref{fig:circuitUs}.
    The global phase factor $\mathrm{e}^{\mathrm{i}\lambda_{\min} x_0}$ is omitted for clarity.}
    \label{fig:circuitZS}
\end{figure*}
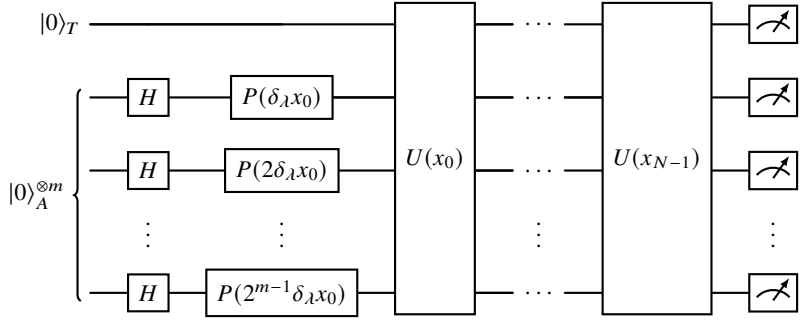

Finally, we assemble the complete quantum circuit by preparing the initial states of both registers.
The ancillary register $A$ is initialized in a uniform superposition using Hadamard gates, which encodes the discretized spectrum of $\lambda$.
The target system is initialized in the state $(\ee^{\ii\lambda x_0}, 0)\T$, as specified in Eq.~(\ref{eq:zsIntegral}), corresponding to the qubit state $|0\rangle_T$ with a phase determined by $\lambda$ and $x_0$.
By substituting the affine mapping in Eq.~\eqref{eq:affine_map}, 
\begin{equation}
    \ee^{\ii \lambda_k x_0} = \ee^{\ii \lambda_{\min} x_0} \prod_{l=0}^{m-1} \exp\big(\ii k_l 2^l \delta_\lambda x_0 \big)
\end{equation}
is decomposed into a product of local single-qubit operations, where $k_l \in \{0, 1\}$ denotes the $l$-th bit in the binary representation of the integer $k = \sum_{l=0}^{m-1} k_l 2^l$. 
Thus, the initial state is deterministically prepared by applying local phase gates $P(2^k \delta_\lambda x_0)$ to each corresponding ancillary qubit, while the global offset $\ee^{i \lambda_{\min} x_0}$ can be tracked classically. 
Following this efficient initialization, the sequential application of the evolution operator $U(x_i)$ completes the fully parallelized circuit architecture for the spatial scattering process. 
The complete quantum circuit for scattering data is shown in Fig.~\ref{fig:circuitZS}.

We next compute the discrete data $\{\lambda_j, c_j\}$ associated with the discrete spectrum.
In principle, one may extract the eigenvalues $\lambda_j$ by casting the Z-S system as a Hamiltonian and applying quantum algorithms such as quantum phase estimation (QPE) or variational quantum algorithm~\cite{Cerezo2021variational}.
However, the finite and typically small number of solitons in physical applications makes a fully quantum approach unnecessary.
Therefore, we adopt the numerical method proposed in Ref.~\cite{cui2022efficient}, which uses Chebyshev polynomials to compute the eigenvalues of the Z-S system and isolate each discrete eigenvalue $\lambda_j$ for a given initial potential $q(x, 0)$.
The corresponding coefficients $c_j$ are then computed from Eq.~\eqref{eq:coefficient}, with $\psi_{l, 1}$ evaluated in the same manner as in Eq.~\eqref{eq:zsIntegral}.

\subsection{Temporal evolution of scattering data}
The temporal evolution of the continuous scattering data in Eq.~\eqref{eq:timeEvolution} is then implemented as a phase operation acting on the target register $T$ in the quantum circuit.
For a fixed $\lambda$, Eq.~\eqref{eq:timeEvolution} is implemented as the diagonal unitary
$\mathrm{diag}(1,\mathrm{e}^{4\mathrm{i}\lambda^2 t})$.
The positive sign in the phase follows from the fact that the direct quantum transform circuit encodes the complex-conjugated scattering data, $a^*(\lambda)$ and $b^*(\lambda)$.
The operation therefore reduces to a single-qubit phase gate $P(4\lambda^2 t)$ acting on the target register $|T\rangle$.

\begin{figure*}
\centering
\begin{quantikz}[row sep=0.35cm, column sep=0.25cm]
    \lstick{$|T\rangle$} & \gate{P(\phi_{0,1})} & \qw ~~\dots~~ & \gate{P(\phi_{m-2,m-1})} & \gate{P(\phi_0)} & \qw ~~\dots~~ & \gate{P(\phi_m)} & \qw & \gate{P(4\lambda_{\min}^2 t)} &  \qw \\
    \lstick[5]{$|A\rangle$} 
    &  \ctrl{-1} & \qw ~~\dots~~ & \qw & \ctrl{-1} & \qw ~~\dots~~ & \qw & \qw & \qw & \qw  \\
    &  \ctrl{-1} & \qw ~~\dots~~ & \qw &  \qw & \qw ~~\dots~~ & \qw & \qw & \qw & \qw  \\
    \setwiretype{n} & & \vdots &  &  & \vdots  \\
    &  \qw & \qw ~~\dots~~ & \ctrl{-4} &  \qw & \qw ~~\dots~~ & \qw & \qw & \qw & \qw  \\
    &  \qw & \qw ~~\dots~~ & \ctrl{-1} &  \qw & \qw ~~\dots~~ & \ctrl{-5} &  \qw & \qw & \qw 
    \end{quantikz}
    \caption{Quantum circuit for the temporal evolution of the continuous scattering data in Eq.~\eqref{eq:temperalOperator}.
    The phase factor $\exp(4\mathrm{i}\lambda_k^2 t)$ is decomposed using the binary representation of the discretized spectral parameter $\lambda_k=\lambda_{\min}+k\delta_\lambda$.
    This decomposition factorizes the temporal evolution into a global phase gate $P(4\lambda_{\min}^2 t)$, singly controlled phase gates $P(\phi_j)$, and doubly controlled phase gates $P(\phi_{j,l})$.
    The resulting circuit requires $\mathcal{O}(m^2)$ controlled phase operations.}
    \label{fig:circuitTime}
\end{figure*}
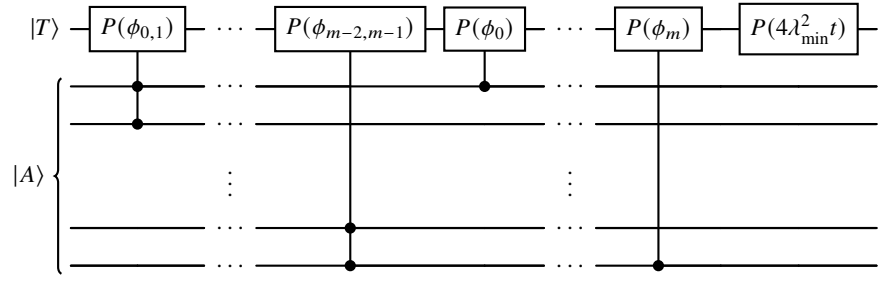

As in the direct scattering transform, we apply this evolution coherently to all discretized spectral values by using the ancillary register $|A\rangle$.
For the grid $\lambda_k=\lambda_{\min}+k\delta_\lambda$, we express the quadratic phase through the binary representation
$k=\sum_{l=0}^{m-1} k_l2^l$, with $k_l\in\{0,1\}$, yielding
\begin{equation}
    \lambda_k^2 = \lambda_{\min}^2 
    + \sum_{j=0}^{m-1} k_j
    \left(2^{j+1}\lambda_{\min}\delta_\lambda+2^{2j}\delta_\lambda^2\right)
    +2\sum_{j,l=0, j<l}^{m-1}k_j k_l 2^{j+l} \delta_\lambda^2.
\end{equation}
Hence the temporal evolution operator
\begin{equation}
    U_{\mathrm{evo}}(t)
    =
    \left( I_A \otimes P(4\lambda_{\min}^2 t) \right)
    \left[ \prod_{j=0}^{m-1} \Lambda_j \big( P(\phi_j) \big) \right]
    \left[ \prod_{j,l=0,j<l}^{m-1} \Lambda_{j,l} \big( P(\phi_{j,l}) \big) \right]
    \label{eq:temperalOperator}
\end{equation} 
factorizes into a global phase gate, single-controlled phase gates, and double-controlled phase gates, with rotation angles $\phi_j = 4t\left(2^{j+1}\lambda_{\min}\delta_\lambda+2^{2j}\delta_\lambda^2\right)$ and $\phi_{j,l} = 2^{j+l+3}t\delta_\lambda^2$.
Here,
\begin{equation}
    \Lambda_{j,l}(V)
    =
    |11\rangle\langle 11|_{j,l}\otimes V
    +
    \left(I-|11\rangle\langle 11|_{j,l}\right)\otimes I_T
\end{equation}
denotes the operation $V$ applied to $|T\rangle$ only when the $j$-th and $l$-th ancillary qubits are both in state $|1\rangle$.
The temporal evolution of the scattering data is therefore implemented with $\mathcal{O}(m^2)$ controlled phase gates, as shown in Fig.~\ref{fig:circuitTime}.

\subsection{Inverse scattering transform}\label{sec:qsvtInverse} 

We then discretize the GLM equation in Eq.~\eqref{eq:glm} into a linear system~\cite{arico2011numerical}.
Among the quantum algorithms for solving the resulting linear system~\cite{harrow2009quantum,childs2017quantum,gilyen2019quantum}, we apply the quantum singular value transformation (QSVT)~\cite{gilyen2019quantum} for solving the discretized GLM equation, because it provides a modular circuit structure and achieves a near-optimal dependence on the condition number $\kappa$ of the coefficient matrix. 

Let the spatial domain be discretized uniformly with step size $h$, yielding grid points $x_j = x_{\min} + jh$ for $j = 0, 1, \dots, N-1$.
To evaluate the integration kernel efficiently, we define the integration variable $z$ and the free collocation variable $y$ on a grid with step size $2h$, such that $y_p = 2ph$ and $z_r = 2rh$ for $p,r = 0, \dots, M$.
Here, $M$ denotes the number of grid intervals, corresponding to $M+1$ collocation points.
To avoid interpolation errors, the spatial and spectral grids must be aligned.
For a truncated spatial domain $x \in [-L, L]$ with step size $h$, the spectral grid for $\lambda$ satisfies
\begin{equation}
    \lambda \in \left[-\frac{\pi}{2h}, \frac{\pi}{2h}\right], \quad \delta_\lambda = \frac{\pi}{2L}. 
    \label{eq:gridAlign}
\end{equation}

For a fixed spatial point $x_j$, we approximate the integral over $z$ using the composite Simpson rule, which yields the diagonal quadrature-weight matrix $D = \mathrm{diag}\{d_r\}_{r=0}^{M}$.
Collocation at $y_p$ reduces the continuous integral system to the linear system
\begin{equation}
    \begin{pmatrix} I & -\Omega_h^* D \\ \Omega_h D & I \end{pmatrix} \begin{pmatrix} B_{h1} \\ B_{h2} \end{pmatrix} = \begin{pmatrix} 0 \\ -\omega_h \end{pmatrix}.
    \label{eq:glmlinearsystem}
\end{equation}
Since $2x_j + y_p + z_r = 2x_{j+p+r}$, $\Omega_h$ is a Hankel matrix with elements $\Omega_{h, pr} = \omega(2x_{j+p+r})$, and $\omega_h$ is a column vector with components $\omega_{h, p} = \omega(2x_{j+p})$.
Solving this system yields the unknown vectors $B_{h1}$ and $B_{h2}$, whose components are $B_{h1, p} = B_1(x_j, y_p)$ and $B_{h2, p} = B_2(x_j, y_p)$, respectively.
Finally, the physical field follows as $q(x_j, t) = 2B_{h2, 0}$. 

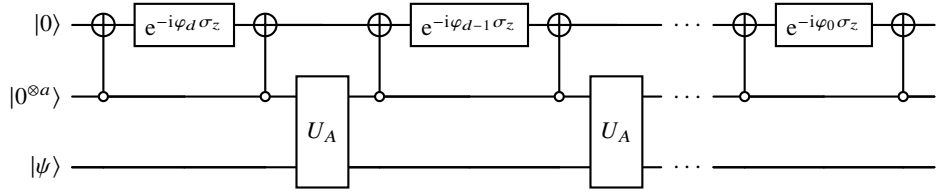
\begin{figure*}
\centering
\begin{quantikz}[row sep=0.4cm, column sep=0.25cm]
            \lstick{$\ket{0}$} & \targ{} & \gate{\ee^{-\mathrm{i}\varphi_d \sigma_z}} & \targ{} & \qw & \targ{} & \gate{\ee^{-\mathrm{i}\varphi_{d-1} \sigma_z}} & \targ{} & \qw & ~~\dots~~ & \targ{} & \gate{\ee^{-\mathrm{i}\varphi_0 \sigma_z}} & \targ{} & \qw \\
            \lstick{$\ket{0^{\otimes a}}$} & \octrl{-1} & \qw & \octrl{-1} & \gate[2]{U_A} & \octrl{-1} & \qw & \octrl{-1} & \gate[2]{U_A} & ~~\dots~~ & \octrl{-1} & \qw & \octrl{-1} & \qw \\
            \lstick{$\ket{\psi}$} & \qw & \qw & \qw &  & \qw & \qw & \qw &  & ~~\dots~~ & \qw & \qw & \qw & \qw
    \end{quantikz}
    \caption{Quantum circuit for implementing a polynomial matrix function $f(A)$ via QSVT.
    Here, $U_A$ is a Hermitian block encoding of $A$.
    The three registers, from top to bottom, are the signal qubit initialized in $\ket{0}$, the $a$-qubit ancilla register initialized in $\ket{0}^{\otimes a}$ for the block encoding $U_A$, and the system register containing the input state $\ket{\psi}$.
    After postselecting the signal and ancilla registers onto $\ket{0}\ket{0}^{\otimes a}$, the state of the system register is proportional to $f(A)\ket{\psi}$.
    Empty dots denote controls conditioned on the ancilla register being in the state $\ket{0}^{\otimes a}$.
    The gates $\ee^{-\mathrm{i}\varphi_j\sigma_z}$ are $Z$ rotations applied to the signal qubit, where $\{\varphi_j\}_{j=0}^d$ are the QSVT phase factors and $d$ is the degree of the polynomial implemented by the QSVT sequence.}
    \label{fig:circuitQSVT}
\end{figure*}

The GLM equation must be solved independently at each spatial grid point $x_j$.
Classically, solving the resulting $N$ independent linear systems is computationally demanding, whereas our quantum approach exploits parallelism by assembling them into a single global block-diagonal system.
This unified large-scale system is then solved simultaneously using QSVT, as illustrated in Fig.~\ref{fig:circuitQSVT}.

To solve the linear system via QSVT, the coefficient matrix $A$ is embedded in a quantum circuit through a block-encoded unitary $U_A$. 
To efficiently construct the block-encoded unitary $U_A$, we decompose the local coefficient matrix in Eq.~\eqref{eq:glmlinearsystem} as $I + \hat{\Omega} \hat{D}$, where $\hat{\Omega} = ((0, \Omega_h)^{\mathrm{T}}, (-\Omega^{*}_h, 0)^{\mathrm{T}})$ and $\hat{D} = \mathrm{diag}(D, D)$.
Since $\hat{\Omega}$ consists exclusively of Hankel blocks, it can be block-encoded with logarithmic gate depth by decomposing it into a linear combination of unitaries (LCU)~\cite{wan2021block}.
The diagonal matrix $\hat{D}$ is directly block-encoded via controlled rotations.
The composite local matrix $I + \hat{\Omega} \hat{D}$ is then constructed using the product rule of block encodings for $\hat{\Omega}\hat{D}$ and the LCU for the addition of $I$.
Leveraging these local encodings, we construct the global block-diagonal system using $\mathcal{O}(\log N)$ ancilla qubits.

With the global coefficient matrix $A$ successfully block-encoded to $U_A$, the QSVT applies a polynomial transformation $P(A)$ to the singular values of $A$.
Solving the linear system requires applying the matrix inverse, which corresponds to the singular-value function $f(x) = 1/x$.
Within QSVT, this inverse function is approximated by a target polynomial $P(x)$.
The circuit realizes this transformation by interleaving the block-encoded unitary $U_A$ with parameterized single-qubit rotations.
The corresponding phase angles are obtained by classical numerical optimization so that the quantum circuit approximates $P(x)$ and outputs the solution state efficiently.

\subsection{Complexity analysis}
We analyze the computational complexity of the proposed quantum algorithm, including both the direct and inverse scattering transforms, and compare it with classical methods and existing quantum methods.
To characterize the discretization resolution, we define the number of spatial grid points $N$, the number of continuous spectral points $N_\lambda = 2^m$, and the number of discrete eigenvalues $N_b$. 
The GLM equation is discretized using $M$ subintervals, yielding $M+1$ collocation points. 

For the quantum direct scattering transform, the initial state preparation applies local phase gates to each auxiliary qubit with a circuit depth $\mathcal{O}(m)$.
The spatial integration then proceeds through sequential applications of the unitary operator $U(x_i)$ at the $N$ spatial grid points.
As shown in Fig.~\ref{fig:circuitUs}, each single-step operator contains two single-qubit rotation gates for the anti-diagonal component $H_{\mathrm{od}}$ and $m$ controlled-$R_z$ gates.
Thus, the gate complexity of the direct transform scales as
\begin{equation}
    \mathcal{C}_{\mathrm{direct}} = \mathcal{O}(N m) = \mathcal{O}(N \log N_{\lambda}).
\end{equation}
By contrast, classically solving the Z-S system for $N_\lambda$ spectral points on a spatial grid of size $N$ requires $\mathcal{O}(N N_\lambda)$ operations with standard integration schemes.

In the present hybrid quantum-classical framework, the discrete eigenvalues $\lambda_j$ and the corresponding coefficients $c_j$ are determined with a classical Chebyshev polynomial method~\cite{cui2022efficient}.
This procedure yields an eigenvalue problem whose matrix size is set by the number of Chebyshev collocation points $N_c$.
In practice, $N_c$ is usually smaller than the original spatial discretization size $N$.
Solving this eigenvalue problem with iterative methods, such as the Lanczos method~\cite{Lanczos1950iteration}, typically requires computational complexity $\mathcal{O}(N_b N_c^2)$.
Future implementations based on variational quantum eigensolver or QPE may reduce the cost of this step.

For the temporal evolution of the scattering data in Eq.~\eqref{eq:temperalOperator}, the circuit contains one global phase gate, $\mathcal{O}(m)$ single-controlled phase gates, and $\mathcal{O}(m^2)$ double-controlled phase gates, as shown in Fig.~\ref{fig:circuitTime}.
Thus, the gate complexity of temporal evolution scales as
\begin{equation}
    \mathcal{C}_{\mathrm{evo}} = \mathcal{O}(m^2) = \mathcal{O}(\log^2 N_\lambda).
\end{equation}
By contrast, classical evolution of all discretized scattering data requires $\mathcal{O}(N_\lambda)$ independent scalar operations.

Extracting the evolved continuous scattering data requires measurements, which introduce a sampling cost dependent on $N_\lambda$ and the target precision $\varepsilon_m$.
By employing quantum amplitude estimation~\cite{Brassard2002_Quantum}, the number of measurement shots required to estimate the corresponding probabilities scales as
\begin{equation}
    M_{\mathrm{meas}} = \mathcal{O}\left(\frac{N_\lambda}{\varepsilon_m} \right).
\end{equation}
Consequently, the total operational complexity for extracting and propagating the continuous scattering data becomes
\begin{equation}
    (\mathcal{C}_{\mathrm{direct}} + \mathcal{C}_{\mathrm{evo}}) M_{\mathrm{meas}} 
    = \mathcal{O}\left( ( N \log N_\lambda + \log^2 N_\lambda) N_\lambda / \varepsilon_m \right).
\end{equation}

Finally, the inverse scattering transform reconstructs the physical field by solving Eq.~\eqref{eq:glmlinearsystem}.
For a single spatial point, Simpson discretization of the integral in Eq.~\eqref{eq:glm} at $M+1$ collocation points yields a linear system of dimension $2(M+1)$.
Classically, one must solve this system independently at all $N$ spatial points, which amounts to solving an aggregated system of dimension $2N(M+1)$.
Here, we assemble these independent equations into a single global block-diagonal matrix and solve the full system simultaneously with QSVT.
Enabled by quantum random access memory~\cite{Giovannetti2008_Quantum}, the block encoding of the local Hankel matrix $\hat{\Omega}$ requires $\mathcal{O}(\mathrm{polylog}(M\chi / \varepsilon))$ operations, where $\chi = \sum | \omega(2x_{j+p+r})|$ is the $L_1$ norm of the displacement matrix and $\varepsilon$ is the target precision.
Meanwhile, the diagonal matrix $\hat{D}$ is block-encoded using $\mathcal{O}(\mathrm{polylog}(M))$ controlled rotations.
Consequently, with the addition of $\mathcal{O}(\log(N))$ auxiliary qubits, the circuit complexity for preparing the solution state scales as
\begin{equation}
    \mathcal{C}_{\mathrm{inverse}} = \mathcal{O}\left( \kappa \chi \, \mathrm{polylog}(NM / \varepsilon) \right),
    \label{eq:QSVTcomplexity}
\end{equation}
where $\kappa$ is the condition number of the global matrix~\cite{wan2021block}.
In contrast, a direct classical implementation must solve a dense linear system at each spatial grid point, with a complexity $\mathcal{O}(NM^3)$.
Therefore, at the level of quantum state preparation, the QSVT formulation offers a possible route to reducing the classical polynomial dependence on $N$ and $M$ to polylogarithmic dependence, under efficient data-access and block-encoding assumptions.

The condition number $\kappa$ and the scaling factor $\chi$ in Eq.~\eqref{eq:QSVTcomplexity} represent worst-case upper bounds on the computational complexity.
In practice, the QSVT complexity is governed by an effective condition number $\kappa_{\mathrm{eff}}$, which is restricted to the active singular values where the target vector has non-negligible projections. 
Numerical experiments demonstrates that $\kappa_{\mathrm{eff}}$ remains well bounded ($\kappa_{\mathrm{eff}} \le 6$) for localized dynamics, such as breathers and two-soliton collisions, without incurring truncation loss. 
For scenarios featuring broader spectral transfer, such as modulational instability, where the condition number naturally grows, classical or quantum preconditioning techniques~\cite{tong2021fast} can be employed to maintain algorithmic efficiency.
A systematic investigation of how the exact bounds of $\chi$ and $\kappa$ behave across diverse nonlinear wave phenomena remains a subject for future study.


In practice, $N_{\lambda}$ typically scales similarly to $N$, $M$ is a prescribed collocation parameter, and our numerical results show that the condition number $\kappa$ remains moderate in the tested cases.
Thus, the overall complexity of our quantum algorithm scales as
\begin{equation}
    \mathcal{C}_{\mathrm{qc}} = (\mathcal{C}_{\mathrm{direct}} + \mathcal{C}_{\mathrm{evo}}) M_{\mathrm{meas}}  + \mathcal{C}_{\mathrm{inverse}} 
    = \mathcal{O}\left(\frac{N N_{\lambda} \log N_{\lambda}}{\varepsilon_m}\right).
    \label{eq:complexity}
\end{equation}
Here, the leading contribution arises from the measurement overhead required to extract the scattering data after the direct transform and temporal evolution.
Note that this asymptotic complexity holds for fault-tolerant quantum computer. 
However, this scaling is not resolved in the subsequent numerical results because classical simulation of the quantum circuit incurs an exponentially growing memory cost. 
The numerical adaptations required to accommodate these constraints are detailed in Sec.~\ref{sec:numericalSetup}.

For comparison, a classical implementation of the Lax framework requires repeated integration of the Z-S system over all spectral points, independent temporal phase updates, and the solution of dense GLM linear systems at all spatial points.
The corresponding total classical complexity scales as
\begin{equation}
    \mathcal{C}_{\mathrm{cc}}
    =
    \mathcal{O}(N N_{\lambda})
    +
    \mathcal{O}(N_{\lambda})
    +
    \mathcal{O}(N M^3)
    =
    \mathcal{O}\left(N(N_{\lambda}+M^3)\right).
\end{equation}
The resulting complexity ratio between the classical and quantum procedures is therefore
\begin{equation}
    \mathcal{S}
    =
    \frac{\mathcal{C}_{\mathrm{cc}}}{\mathcal{C}_{\mathrm{qc}}}
    =
    \mathcal{O}\left(
    \frac{\varepsilon_m M^3}
    {N_\lambda \log N_{\lambda}}
    \right).
\end{equation}
A practical quantum advantage is therefore expected when the inverse GLM reconstruction dominates the classical cost, namely in the regime $M^3 \gg N_{\lambda}\log N_{\lambda}/\varepsilon_m$, or when the algorithm estimates global observables directly from quantum states without reconstructing all scattering data.
Under efficient data-access and block-encoding assumptions, and for moderate $\kappa$, the QSVT-based inverse transform provides the main source of possible asymptotic acceleration.

Moreover, we compare the proposed quantum algorithm with existing quantum-computing approaches for solving the NLSE.
Most existing frameworks rely on the split-step Fourier method (SSFM), which separates the time evolution into linear and nonlinear steps.
Typically, the linear step is implemented with the quantum Fourier transform, whereas the nonlinear step is treated with different strategies.
The nonlinear term is implemented by a parameterized quantum circuit in Ref.~\cite{kocher2025numerical}, with a time complexity $\mathcal{O}( N_t ( d \log N + \log^2 N )/ \varepsilon^{2} )$, where $N_t$ denotes the number of time steps, $d$ is the depth of the parameterized quantum circuit, and $\varepsilon$ denotes the measurement precision of the cost function.
However, this estimate does not include the classical training cost, which may introduce substantial additional overhead.
Alternatively, the nonlinear term can be reconstructed by measuring the Fourier components of low-wavenumber modes, which reduces the measurement overhead~\cite{weng2026quantum}.
The overall complexity of this hybrid framework scales as $\mathcal{O}(N_t^2 M_f \log^2 N / \varepsilon^2)$, where $M_f$ denotes the number of retaining Fourier modes.

In contrast, the proposed algorithm performs the evolution entirely in the spectral domain and avoids iterative time stepping.
Consequently, the overall computational complexity is independent of the evolution time.
This feature gives our approach favorable scaling with respect to the simulated evolution time compared with existing methods, especially for long-time simulations where the number of time steps $N_t$ becomes very large.

\section{Numerical results}\label{sec:result}

\subsection{Numerical setup}\label{sec:numericalSetup}

We assess the robustness and capabilities of our method through four representative numerical experiments.
These tests comprise a Gaussian wave packet subject to quantum noise in the direct-transform circuit, two-soliton collisions that test nonlinear interactions, breather solutions that probe strongly localized nonlinear wave dynamics, and modulational instability (MI) that examines dynamically evolving multiscale nonlinear processes.
The QSVT implementation uses an open-source library~\cite{qsvtGithub}.

Because of runtime and memory limitations in the quantum emulator implemented on a classical computer, we use several practical trade-offs.
First, to accelerate simulations of the direct scattering transform, we explicitly construct the gate matrices in a dense-matrix representation and multiply the corresponding unitary operators directly, rather than decomposing them into elementary gates.
To reduce the severe memory cost associated with the high dimensionality of the global matrix in the QSVT-based inverse scattering transform, we solve the linear system in Eq.~\eqref{eq:glmlinearsystem} independently at each spatial grid point.
Besides, all quantum results are extracted from the statevector.
By eliminating statistical shot noise, this procedure provides a noise-free baseline that isolates the effects of the imposed noise model.
Although these practical numerical adaptations are dictated by hardware limitations, they preserve the mathematical fidelity of the operations and do not affect the verification of the algorithmic components.
Consequently, these numerical results demonstrate the mathematical consistency and noise robustness of the proposed quantum framework.
The source code of this study is available at https://github.com/YYgroup/QuantumLaxPair.

\subsection{Gaussian wave packet under quantum noise}

We use a Gaussian wave packet, $q(x)=\mathrm{e}^{-x^2/2}/10$, as a benchmark for the quantum direct-scattering circuit.
The purpose of this test is to quantify how quantum noise in the forward transform affects the extracted scattering data and the subsequent reconstruction.
The wave packet is first encoded into the noisy direct-scattering circuit, from which the scattering data are extracted.
We then reconstruct the wave packet from these data using the inverse scattering transform, without applying any time-evolution step in the spectral domain.
This setup removes the dynamical evolution from the pipeline and therefore isolates the intrinsic accuracy and noise robustness of the direct-scattering circuit.

The computations are performed on the physical domain $x\in[-6,6]$.
The spectral domain is $\lambda\in[-26.049,26.049]$, with its range determined by Eq.~\eqref{eq:gridAlign}.
Both domains are uniformly discretized into 200 grid points.
For the inverse reconstruction, the linear system in Eq.~\eqref{eq:glmlinearsystem} is assembled at each physical grid point using 16 collocation points, namely $M=15$.

To simulate stochastic hardware noise, we apply a depolarizing error to each spatial integration gate $U(s_j)$ with probability $p$.
We sample 20,000 independent noise realizations using a statevector simulator.
For each run, we extract the exact final statevector for the corresponding noise trajectory.
The resulting scattering data $a(\lambda)$ and $b(\lambda)$ are averaged over the noise ensemble and then used to reconstruct the physical field.
We test the circuit at noise levels $p = 10^{-5}$, $10^{-4}$, $10^{-3}$, and $5 \times 10^{-3}$.

\begin{figure*}
    \centering
    \includegraphics[width=\linewidth]{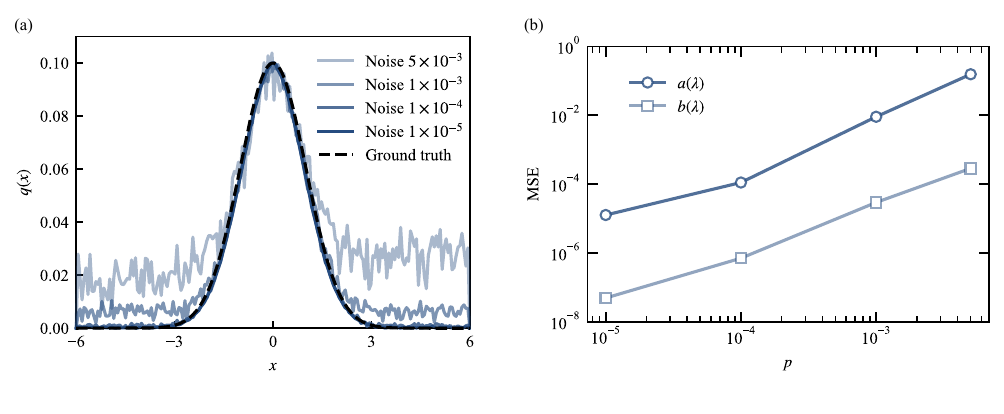}
    \caption{Effect of quantum noise on the direct transform circuit.
    (a) Reconstructed physical field $q(x)$ for different depolarizing error probabilities $p$.
    The reconstruction remains close to the ground truth for $p \le 10^{-3}$, indicating moderate robustness of the direct transform circuit.
    (b) Mean-squared errors of the scattering data $a(\lambda)$ and $b(\lambda)$ as functions of the noise probability $p$.
    To remove the ambiguity from arbitrary global phases introduced by the quantum circuit, the errors are computed from the absolute values of the scattering data and compared with numerical benchmarks obtained by the DOP853 method~\cite{dormand1986reconsideration}.
    The error trends show that the circuit preserves acceptable accuracy for noise levels up to $p = 10^{-3}$.}
    \label{fig:noise}
\end{figure*}

Figure~\ref{fig:noise}(a) shows the recovered physical field, and Fig.~\ref{fig:noise}(b) shows the mean squared error (MSE) of the scattering data.
Overall, the circuit exhibits moderate robustness against quantum noise.
The recovered physical field remains accurate for error probabilities up to $p = 10^{-3}$, with the MSE of the scattering data staying below $10^{-2}$.
Current quantum hardware, particularly trapped-ion and superconducting platforms, achieves single-qubit error rates below $10^{-3}$ and two-qubit error rates approaching the $10^{-3}$ threshold~\cite{vukvsic2026comparative}.
The present results therefore motivate further tests on near-term devices.

\subsection{Soliton collisions}
We test our method on the collision of two solitons to demonstrate its ability to capture nonlinear dynamics.
In Eq.~\eqref{eq:nls}, the nonlinear term $|q|^2$ generates an attractive interaction between the solitary waves during the collision.
This interaction produces a clear trajectory shift for both solitons~\cite{shabat1972exact}.
Because soliton collisions are elastic, the amplitude, width, and velocity of each soliton are conserved asymptotically before and after the interaction.
Accurate reproduction of these features provides a stringent benchmark for our quantum framework's capacity to simulate nonlinear dynamics.

For the numerical setup, we use the initial condition
\begin{equation}
    q(x, t=0) 
    = -3\ii \operatorname{sech}\left( -12 - 3 x \right) \ee^{-0.8\ii x} 
    - 2\ii \operatorname{sech}\left( 7.5 - 2 x \right) \ee^{\ii x/2},
\end{equation}
which represents a superposition of two distinct solitons prepared for a head-on collision.
The first term corresponds to a high-amplitude and narrow soliton initially centered at $x = -4$ and propagating in the positive $x$-direction. 
The second term describes a lower-amplitude and broader soliton initially centered at $x = 3.75$ and propagating in the negative $x$-direction. 
We discretize the physical domain $x \in [-15, 15]$ and the corresponding spectral domain $\lambda \in [-13.351, 13.351]$ using a uniform grid with 256 points.
The spectral domain is determined by Eq.~\eqref{eq:gridAlign}.
During the inverse reconstruction, we construct the linear system \eqref{eq:glmlinearsystem} by selecting a subset of 16 collocation points ($M = 15$) for each local physical point.
We evolve the system from $t = 0$ to $t = 8$ and record the results at intervals of $\delta_t = 0.08$.
For comparison, we also solve the equation using the classical SSFM with a time step of $\delta_t = 0.001$ to ensure numerical stability.

\begin{figure}
    \centering
    \includegraphics[width=0.5\linewidth]{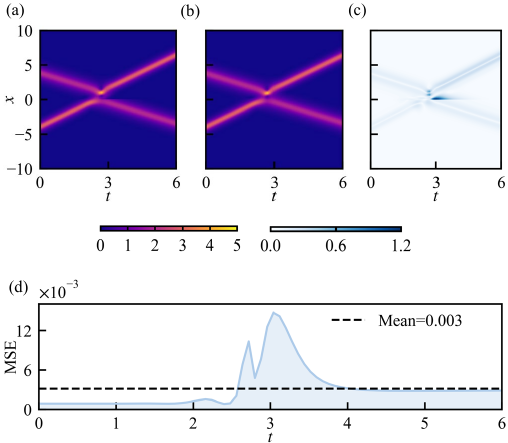}
    \caption{Simulation results of two-soliton collision: (a) $|q|$ obtained by the present method, (b) $|q_{\mathrm{SSFM}}|$ obtained by the SSFM, and (c) the absolute error $||q| - |q_{\mathrm{SSFM}}||$. 
    (d) Time evolution of the MSE.}
    \label{fig:soliton}
\end{figure}

Figure~\ref{fig:soliton}(a) shows the time evolution of the two solitons computed by our quantum framework, while Fig.~\ref{fig:soliton}(b) shows the corresponding SSFM result for comparison.
For visual clarity, the figure displays only the window $x\in [-10, 10]$ and $t \in [0, 6]$.
For quantitative comparison, Fig.~\ref{fig:soliton}(c) shows the pointwise error between $|q|$ and $|q_{\mathrm{SSFM}}|$, with the temporal evolution of the MSE in Fig.~\ref{fig:soliton}(d).
The MSE remains small during most of the evolution, except for a transient increase near the collision.
This agreement confirms that our quantum framework resolves the nonlinear dynamics of soliton interactions.

\subsection{Breather}
As a more demanding validation of our method, we investigate breather dynamics, which provide a representative class of coherent structures in the focusing NLSE~\cite{dudley2014instabilities}.
Breathers are coherent wave structures with localized amplitude modulations that undergo periodic cycles of spatial concentration and dispersion during propagation.
They describe localized energy-focusing events in a wide range of dispersive media.
In deep-water hydrodynamics, breather solutions provide a theoretical model for rogue waves and describe highly localized transient surface elevations generated by nonlinear wave interactions~\cite{osborne2019highly}.
In 2D electron plasmas, similar structures govern intense spatiotemporally localized charge-density excitations that arise from the modulational instability of continuous plane-wave backgrounds~\cite{zabolotnykh2023nonlinear}.
A breather solution of the NLSE is
\begin{equation}
    q(x, t) = \frac{4 \ee^{\ii t} \left[\cosh(3x) + 3 \ee^{8\ii x} \cosh(x) \right]}{\cosh(4x) + 4\cosh(2x) + 3 \cos(8t)}.
    \label{eq:breather}
\end{equation}

We set the physical domain $x\in[-10, 10]$ and the corresponding spectral domain $\lambda \in [-20.028, 20.028]$ determined by Eq.~\eqref{eq:gridAlign}.
Both domains are discretized using 256 grid points.
We use 16 collocation points to construct Eq.~\eqref{eq:glmlinearsystem}.
The simulation runs from $t=0$ to $t=2\pi$.
The time step is $\delta_t=0.02\pi$, which resolves the periodic dynamics.

\begin{figure*}
    \centering
    \includegraphics[width=\linewidth]{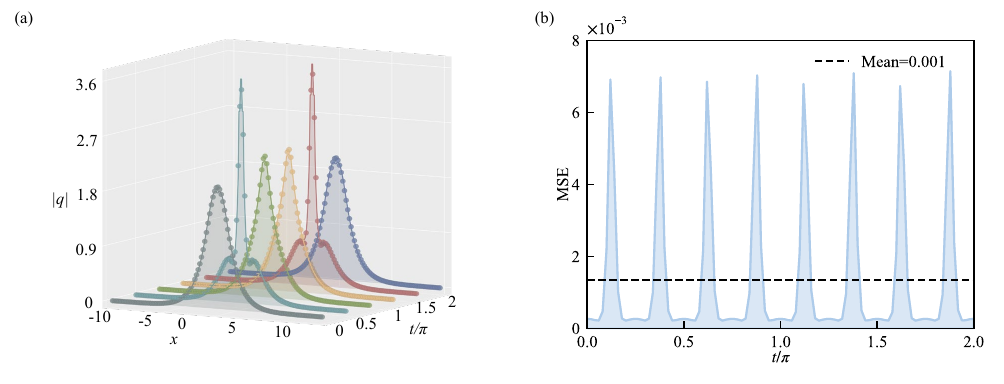}
    \caption{Simulation results for the breather.
    (a) Comparison between $|q|$ obtained by the present method and the exact solution.
    (b) Time evolution of the MSE, with the solid black line indicating the overall average MSE of $0.001$.}
    \label{fig:breather}
\end{figure*}

Figure~\ref{fig:breather}(a) compares the absolute value $|q|$ obtained by the present method with the exact solution in Eq.~\eqref{eq:breather}.
The two profiles overlap almost completely over the entire computational domain, which demonstrates that the method accurately captures the spatiotemporal structure of the breather dynamics.
Quantitatively, Fig.~\ref{fig:breather}(b) shows the MSE, with an average value near $0.001$, indicating high numerical accuracy.

\subsection{Modulational instability}
In more realistic systems, beyond soliton collisions and breather dynamics, wave dynamics often originate from the instability of continuous background states and then evolve into broadband multiscale structures. 
For example, the modulation instability (MI) poses a substantially greater challenge because nonlinear amplification, spectral transfer, and dynamically emerging localized wave packets occur simultaneously~\cite{solli2007optical, chen2024drift}.
In MI, a weak periodic perturbation on an initially uniform background wave is rapidly amplified by the nonlinear interaction in the NLSE. 
As the system evolves, the background breaks into strongly modulated localized wave structures accompanied by spectral transfer.
Physically, MI plays a central role in rogue-wave generation, optical filamentation, and wave-turbulence formation~\cite{dudley2014instabilities}.

We therefore simulate MI to further assess the robustness of our quantum framework for dynamically evolving multiscale nonlinear phenomena.
The initial condition is
\begin{equation}
    q(x, 0) = \ee^{-(|x| / 10)^8} \left(1 + 0.02 \cos (1.2 x)\right) ,
\end{equation}
where the finite-width wave packet serves as the background wave and enforces compatibility with the zero boundary condition.
We multiply this background by a weak cosine perturbation to trigger the instability.
The simulation uses the physical domain $x\in[-24, 24]$ and the corresponding spectral domain $\lambda \in [-16.722, 16.722]$.
Both domains are discretized using 512 grid points.
In the inverse scattering transform, we use 64 collocation points to construct Eq.~\eqref{eq:glmlinearsystem}.
The simulation runs from $t = 0$ to $t = 10$, and the results are recorded at intervals of $\delta_t = 0.1$.
Furthermore, to accommodate memory constraints, we simplify the direct scattering calculation into independent integrations for each $\lambda$ and apply a singular-value-based preconditioning step prior to the QSVT solver.
Specifically, we construct a preconditioning matrix to artificially raise singular values below a prescribed threshold, thereby reducing the condition number.

\begin{figure}
    \centering
    \includegraphics[width=0.5\linewidth]{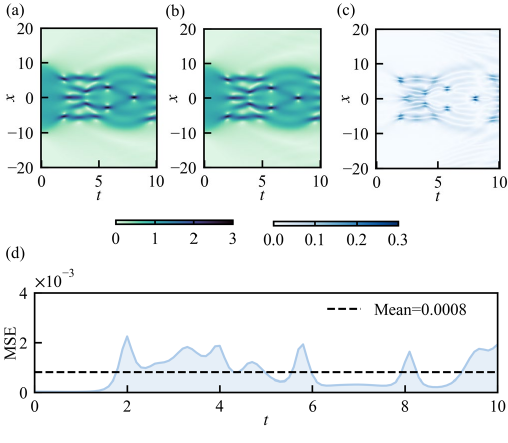}
    \caption{Simulation results for the modulational instability: (a) $|q|$ obtained by the present method, (b) $|q_{\mathrm{SSFM}}|$ obtained by the SSFM, and (c) the absolute error $||q| - |q_{\mathrm{SSFM}}||$.
    (d) Time evolution of the MSE, with the black dashed line indicating the overall average MSE of 0.0008. The consistently low MSE throughout the MI evolution confirms the accuracy of the proposed framework.}
    \label{fig:MI}
\end{figure}

Figure~\ref{fig:MI}(a) shows the MI dynamics computed by our quantum framework, while Fig.~\ref{fig:MI}(b) shows the corresponding SSFM result for comparison.
For visual clarity, only the region $x\in[-20,20]$ is shown.
For quantitative evaluation, Fig.~\ref{fig:MI}(c) shows the absolute error distribution, and Fig.~\ref{fig:MI}(d) shows the temporal evolution of the MSE.
The nearly identical profiles in Figs.~\ref{fig:MI}(a) and~\ref{fig:MI}(b), together with the low average MSE of $8\times10^{-4}$ over the full evolution, demonstrate close agreement between our method and the classical SSFM benchmark.
These results further confirm that the proposed framework captures dynamically evolving multiscale nonlinear dynamics governed by the NLSE.

\section{Conclusions}
\label{sec:conclusion} 
We propose a quantum computing framework for solving the 1D NLSE via the Lax method.
Specifically, we map the Z-S system to a highly parallel quantum circuit that executes the direct transform to extract the scattering data, implement decoupled linear evolution in spectral space, and use QSVT for the inverse scattering transform.
This semi-analytical approach maps coupled nonlinear dynamics to decoupled linear evolution in spectral space, thereby resolving nonlinear evolution equations within a unitary quantum computing framework.

The quantum implementation of this semi-analytical framework rests on efficient realizations of both the direct and inverse scattering transforms.
For the direct scattering transform, we construct a coherent spectral-parallel circuit, which reduces the circuit-depth dependence on the number of spectral points.
For the inverse scattering transform, we formulate the discretized GLM equation as a QSVT-compatible linear-system problem, providing a route for efficient preparation of the reconstructed solution state.
These subroutines achieve polylogarithmic circuit depth reduction on fault-tolerant quantum devices.
Furthermore, since the temporal propagation is performed analytically in the scattering domain, the computational cost of this framework is independent of the number of time steps. 
This feature distinguishes the present approach from time-marching quantum algorithms for the NLSE~\cite{kocher2025numerical, weng2026quantum}, offering a clear advantage for long-time simulations.
However, numerical errors in the inverse transform can accumulate over time due to the highly oscillatory nature of the scattering phase. 
The experimental validation of asymptotic speedups and a rigorous error analysis of long-time simulations, particularly regarding the dynamical evolution of the scaling factor $\chi$ and the condition number $\kappa$, are left for future work.

We validate the robustness and fidelity of the proposed framework by simulating Gaussian wave packets under quantum noise, two-soliton collisions, breather dynamics, and modulational instability using a quantum emulator implemented on a classical computer. 
The results demonstrate that the method accurately resolves complex nonlinear interactions and multiscale spatio-temporal phenomena.
Furthermore, our quantum direct scattering circuit demonstrates moderate robustness to depolarizing noise, maintaining acceptable reconstruction fidelity for gate error rates up to $10^{-3}$, comparable to error rates reported for current platforms, although full-device implementation requires a separate resource analysis.

However, the proposed method has two major limitations.
First, owing to its intrinsic reliance on the Lax formalism, the framework is restricted to integrable systems.
Moreover, the quantum-circuit design for the direct transform is tailored to the specific mathematical structure of the corresponding Lax pair.
Extending the framework to other equations therefore requires custom circuit constructions and operator encodings, potentially increasing both implementation complexity and circuit depth.
Second, the current workflow is not fully end-to-end.
It requires intermediate measurements to extract the scattering data prior to the inverse transform, thereby partially compromising the overall quantum speedup.

Looking forward, the principles underlying this framework can be generalized beyond the NLSE to a broader class of integrable nonlinear systems governed by the zero-curvature condition.
This offers a scalable pathway for the quantum simulation of highly nonlinear physical phenomena, such as deep-water waves and quantum turbulence.
Future developments will focus on integrating fully quantum eigenvalue-extraction methods, such as QPE, and optimizing block-encoding strategies to further reduce sampling overhead.
These developments could enable a fully coherent quantum workflow and clarify the requirements for establishing an end-to-end quantum advantage.

\section*{Acknowledgments}
We thank Hao Su for helpful discussions. This work has been supported in part by the National Natural Science Foundation of China (grant Nos.~12525201, 12432010, 12588201, 12532013, and 12588301). 

\section*{Data availability}
The data analysis and numerical simulation codes for this study are available for download at \url{https://github.com/YYgroup/QuantumLaxPair}.

\appendix

\section{Derivation of the coefficient $c_j$}
\label{sec:coeffDerivation}
We detail the derivation of Eq.~\eqref{eq:coefficient}. 
The second column of Eq.~\eqref{eq:scatterMatrix} reads
\begin{equation}
    \psi_{r,2} = -b(\lambda) \psi_{l, 1} + a(\lambda) \psi_{l, 2}.
\end{equation}
For two arbitrary vectors $a = (a_1, a_2)^\mathrm{T}$ and $b = (b_1, b_2)^\mathrm{T}$, define the Wronskian $W(a,b)=a_1b_2-a_2b_1$.
Since the Wronskian is a bilinear function, applying it to $\psi_{l, 1}$ and $\psi_{r, 2}$ yields
\begin{equation}
        W(\psi_{l, 1}, \psi_{r, 2}) = -b(\lambda) W(\psi_{l, 1}, \psi_{l, 1}) + a(\lambda) W(\psi_{l, 1}, \psi_{l, 2})
        = a(\lambda) \mathrm{det}( \varPsi_{l}).
\end{equation}
The Liouville's formula~\cite{teschl2012ordinary} $\p \mathrm{det}( \varPsi_{l}) / \p x = \operatorname{tr}(U) \mathrm{det}( \varPsi_{l}) =  0$ 
implies $\mathrm{det}( \varPsi_{l}) = \lim_{x \to -\infty} \mathrm{det}( \varPsi_{l}) = 1$. 
Consequently, 
\begin{equation}
    a(\lambda) = W(\psi_{l, 1}, \psi_{r, 2})
    \label{eq:aWronskian}
\end{equation}
is expressed in terms of the Wronskian of two Jost solutions.
Differentiating Eq.~\eqref{eq:aWronskian} with respect to $\lambda$ yields
\begin{equation}
    a'(\lambda) = W\left(\partial_\lambda \psi_{l, 1} ,~\psi_{r, 2}\right) + W\left(\psi_{l ,1},~ \partial_{\lambda}\psi_{r, 2}\right).
    \label{eq:aDiff}
\end{equation}

To evaluate $a'(\lambda_j)$ using Eq.~\eqref{eq:aDiff}, we investigate the asymptotic behavior of the Jost solutions at the discrete eigenvalues $\lambda_j$. According to Eq.~\eqref{eq:scatterMatrix}, $\psi_{l, 1}$ and $\psi_{r, 2}$ are related by
\begin{equation}
    \psi_{l, 1} = a(\lambda) \psi_{r, 1} - b^*(\lambda) \psi_{r, 2}.
\end{equation}
Since $a(\lambda_j) = 0$, the asymptotic behaviors of $\partial_\lambda \psi_{l, 1}$ and $\psi_{r, 2}$ as $x \to +\infty$, evaluated at $\lambda = \lambda_j$, are expressed as
\begin{equation}
          \lim_{x \to +\infty} \partial_\lambda \psi_{l, 1}(x, \lambda) \big|_{\lambda_j} =  \begin{pmatrix}
                \ii x \ee ^{\ii \lambda_j x } \\ 0
          \end{pmatrix}, \quad 
        \lim_{x \to +\infty} \psi_{r, 2}(x, \lambda_j) = b(\lambda_j) \begin{pmatrix}
            \ee^{\ii \lambda_j x} \\ 0
        \end{pmatrix} .
\end{equation}
Consequently, we obtain
\begin{equation}
    \lim_{x\to +\infty} W\left(\partial_\lambda \psi_{l, 1}(x, \lambda) \big|_{\lambda_j} ,~\psi_{r, 2}(x, \lambda_j) \right) = 0. 
    \label{eq:WPosInf}
\end{equation}
Similarly, as $x \to -\infty$, the asymptotic behavior at $\lambda_j$ becomes
\begin{equation}
          \lim_{x \to -\infty} \partial_\lambda \psi_{l, 1}(x, \lambda) \big|_{\lambda_j} =  \begin{pmatrix}
                a'(\lambda_j) \ee^{\ii \lambda_j x} \\ b^{*}(\lambda_j) \ii x \ee^{-\ii \lambda_j x}  - b^{'*} (\lambda_j) \ee^{-\ii \lambda_j x } 
          \end{pmatrix}, \quad 
        \lim_{x \to -\infty} \psi_{r, 2}(x, \lambda_j) = \begin{pmatrix}
            0 \\ \ee^{-\ii \lambda_j x}
        \end{pmatrix} ,
\end{equation}
which yields
\begin{equation}
    \lim_{x\to -\infty} W\left(\partial_\lambda \psi_{l, 1}(x, \lambda) \big|_{\lambda_j} ,~\psi_{r, 2}(x, \lambda_j) \right) = a'(\lambda_j).
    \label{eq:WNegInf}
\end{equation}
By combining Eq.~\eqref{eq:WPosInf} and Eq.~\eqref{eq:WNegInf}, we obtain
\begin{equation}
    a'(\lambda_j) =  -\int_{-\infty}^{+\infty} \partial_x W\left(\partial_\lambda \psi_{l, 1}(x, \lambda) \big|_{\lambda_j} ,~\psi_{r, 2}(x, \lambda_j) \right) \dif x .
    \label{eq:aDiffInt}
\end{equation}

Next, we rewrite the integral in Eq.~\eqref{eq:aDiffInt} in terms of $\psi_l$. From Eq.~\eqref{eq:laxtime}, any solution $\psi = \left(\psi^{(1)}, \psi^{(2)}\right)^{\mathrm{T}}$ of the Z-S system satisfies
\begin{subequations}
    \begin{empheq}[left=\empheqlbrace]{align}
        \partial_x \psi^{(1)} &= \ii \lambda \psi^{(1)} - q(x) \psi^{(2)}  \\
        \partial_x \psi^{(2)} &= -\ii \lambda \psi^{(2)} + q^*(x) \psi^{(1)}.
    \end{empheq}
    \label{eq:LaxTimePart}
\end{subequations}
Differentiating Eq.~\eqref{eq:LaxTimePart} with respect to $\lambda$ yields
\begin{subequations}
    \begin{empheq}[left=\empheqlbrace]{align}
        \partial_{x \lambda} \psi^{(1)} &= \ii \psi^{(1)} + \ii \lambda \partial_{\lambda} \psi^{(1)} - q(x) \partial_{\lambda} \psi^{(2)}  \\
        \partial_{x \lambda} \psi^{(2)} &= -\ii \psi^{(2)} -\ii \lambda \partial_{\lambda} \psi^{(2)} + q^*(x) \partial_{\lambda} \psi^{(1)}.
    \end{empheq}
    \label{eq:LaxTimePartDif}
\end{subequations}
Substituting Eq.~\eqref{eq:LaxTimePart} and Eq.~\eqref{eq:LaxTimePartDif} into the spatial derivative of the Wronskian yields
\begin{equation}
    \partial_x W\left(\partial_\lambda \psi_{l, 1}(x, \lambda) \big|_{\lambda_j} ,~\psi_{r, 2}(x, \lambda_j) \right) = \ii \left(\psi^{(1)}_{l, 1} \psi^{(2)}_{r, 2} + \psi^{(2)}_{l, 1} \psi^{(1)}_{r, 2}\right).
\end{equation}
At the discrete eigenvalue $\lambda_j$, the relation $\psi_{r, 2} = b(\lambda_j) \psi_{l, 1}$ gives the identity
\begin{equation}
    a'(\lambda_j) = -2\ii b(\lambda_j) \int_{-\infty}^{+\infty}\psi^{(1)}_{l, 1} \psi^{(2)}_{l, 1}  \dif x,
    \label{eq:aDiffIntFinal}
\end{equation}
which is expressed entirely in terms of the components of $\psi_{l, 1}$.
Finally, substituting Eq.~\eqref{eq:aDiffIntFinal} into $c_j = -\ii b(\lambda_j) / a'(\lambda_j)$ yields Eq.~\eqref{eq:coefficient}.

\bibliographystyle{elsarticle-num-names}
\bibliography{ref}

\end{document}